\documentstyle[12pt]{article}

\newcommand{\bea}{\begin{eqnarray}}
\newcommand{\eea}{\end{eqnarray}}
\newcommand{\co}{\; \; ,}
\newcommand{\per}{\; \; .}
\newcommand{\nn}{\nonumber \\}

\newcommand{\be}{\begin{equation}}
\newcommand{\ee}{\end{equation}}
\newcommand{\ba}{\begin{array}{c}}
\newcommand{\ea}{\end{array}}
\newcommand{\dis}{\displaystyle}
\newcommand{\bd}{\begin{displaymath}}
\newcommand{\ed}{\end{displaymath}}
\newcommand{\br}{\be\renewcommand{\arraystretch}{1.3}\begin{array}{l}}
\newcommand{\er}{\ea \renewcommand{\arraystretch}{1}\ee}
\newcommand{\bo}{\bd\renewcommand{\arraystretch}{1.3}\begin{array}{l}}
\newcommand{\eo}{\ea \renewcommand{\arraystretch}{1}\ed}

\newcommand{\apbc}{(\alpha + \beta)^C}
\newcommand{\ambc}{(\alpha - \beta)^C}
\newcommand{\apbn}{(\alpha + \beta)^N}

\newcommand{\ambcn}{(\alpha - \beta)^{C,N}}
\newcommand{\apbcn}{(\alpha + \beta)^{C,N}}

\begin{document}
\begin{titlepage} \vspace{0.2in} \begin{flushright} BUTP-94/26 \\
LNF-94/037 (P) \\  \end{flushright} \vspace*{1.5cm} \begin{center} {\LARGE
\bf  $\gamma \gamma \to \pi^0 \pi^0$ and $\eta \to \pi^0 \gamma \gamma$
at low energy \\within the \\Extended Nambu Jona-Lasinio Model$^*$\\}
\vspace*{0.8cm} {\bf S. Bellucci$^a$ and C. Bruno$^b$}\\\vspace*{0.8cm}

\noindent INFN-Laboratori Nazionali di Frascati \\
P.O.Box 13, I-00044
Frascati, Italy

\vspace*{1.8cm} {\bf   Abstract}  \\
\end{center}
{Within the Extended Nambu
Jona-Lasinio model, we calculate the $O(p^6)$ counterterms entering the
low-energy
expansion of the $\gamma \gamma \to \pi^0 \pi^0$ and the $\eta \to \pi^0
\gamma \gamma$ amplitudes in Chiral Perturbation
Theory. For $\gamma \gamma \to
\pi^0 \pi^0$ our results are compatible with both the experimental data and the
two-loop calculation using meson resonance saturation.
For the $\eta$ decay we find
$\Gamma(\eta\to\pi^0 \gamma\gamma)=0.58\pm 0.3$ eV which is in
agreement with experiment within one standard deviation.
We also give predictions
for the neutral pion polarizabilities and compare
them with the results obtained from resonance saturation.}

\vfill

{\small
\noindent *) Work supported in part by the EEC Human Capital and Mobility
Program

\noindent a) e-mail: bellucci@lnf.infn.it

\noindent b) also at: Institute for Theoretical Physics, University of Bern,
Sidlerstrasse 5, CH-3012 Bern, Switzerland; e-mail: cbruno@butp.unibe.ch}

\end{titlepage}

\section{\bf Introduction}

Over the last few years the transitions $\gamma \gamma \to \pi^0 \pi^0$
and $\eta \to \pi^0 \gamma \gamma$ have
been studied in the framework of Chiral Perturbation Theory ($\chi$PT).
These two processes have in common the fact that their amplitude starts
at $O(p^4)$ and that there is no contribution from tree diagrams at this
lowest order.

The first transition has recently been analysed in \cite{bgs} at the
two-loop level. The $O(p^6)$ counterterms have been evaluated assuming
resonance saturation and it is shown that the predicted cross-section fits
well the existing data \cite{crystal}. It has been noted by
several authors \cite{ko1}-\cite{bhn}
that the contribution of these counterterms to
the cross-section is very small in the threshold region. On
the other hand, when not only
high-precision data from DA$\Phi$NE become available for energies up to
$600$ MeV, but also the theoretical predictions are improved by unitarization
procedure (this requires calculating the full $\gamma \gamma \to \pi^+ \pi^-$
amplitude to two-loops), it may become possible to extract
one linear combination of the $O(p^6)$ coupling constants.

Concerning $\eta \to \pi^0 \gamma \gamma$ a particular feature of this
decay is that the $O(p^4)$ one-loop contribution to the width is more than two
orders of magnitude smaller than the experimental measurement. In \cite{abbc}
this suppression has been explained in physical terms and a partial analysis
of higher loop contributions has been carried out. It is expected that the
contribution of the $O(p^6)$ counterterms should
account for a large part of the
amplitude. The analysis of \cite{abbc} gives a reasonable estimate of the order
of magnitude for the decay width. However the latter is still a factor two
smaller
than the experimental value \cite{pdb}
$\Gamma(\eta\to\pi^0\gamma\gamma)=0.85\pm
0.18$ eV. New data may become available in the foreseeable future at SATURNE in
Saclay, where a proposal for measuring this decay width has been recently
approved.

The S-matrix elements for both transitions depend on the same set of $O(p^6)$
coupling constants. Throughout the following we restrict ourselves to
the large $N_c$-limit, where only three low-energy couplings enter these
transitions.
The theoretical analysis is not yet refined enough and the present
experiments are not sufficiently precise, to extract the
value of
these  coupling constants from the data. Therefore it is important
to make predictions for these couplings (this is particularly relevant in
the case of the $\eta\to \pi^0 \gamma \gamma$ decay width which has a
strong dependence on the counterterms). One possibility is to
estimate their value using the resonance saturation method, which
provides successful predictions to $O(p^4)$ \cite{egpr,eglpr}.
However it is not known whether the $O(p^6)$ couplings are actually
saturated by resonance exchange. In addition to the estimate based
on resonance
saturation, one can find in the literature a calculation of two of the
$O(p^6)$ coupling constants using the
chiral quark model \cite{bdv}.

The purpose of our paper is to compute the three $O(p^6)$
coupling constants of the large $N_c$-limit in the framework of the
Extended Nambu Jona-Lasinio (ENJL) model. Within this model the low-energy
effective action of QCD has been derived \cite{bbr} to $O(p^4)$, as well
as some of the coupling constants governing the low-energy behaviour of the
lowest spin-1 and spin-0  resonances
\cite{bbr,prades}. The model has only three parameters (which is an advantage
compared to the resonance saturation method where the number of parameters
increases when new transitions and higher chiral order are considered) and the
agreement with experiment is good.

This paper is organized as follows.
In section 2 we present the structure of
the $O(p^6)$ chiral Lagrangian in the large $N_c$-limit. Section 3 contains
the expressions for the amplitudes of the transitions and the pion
polarizabilities. Section 4 is devoted to the ENJL predictions
for the $O(p^6)$ coupling constants. Finally in section 5 we discuss the
numerical results and compare our predictions with those of the resonance
saturation
method and with the existing data.

\section{\bf  $O(p^6)$ chiral Lagrangian}

Let us collect as usual the octet of Goldstone bosons in the unitary
unimodular matrix $U$

\be\label{19}
U = \hbox{exp} \left( -i \sqrt 2 {\Phi_8(x) +\Phi_1(x) \over f_0} \right),
\ee

\vskip 0.4cm

\noindent where $f_0 \simeq f_{\pi} = 93.2$ MeV and
($\stackrel{\rightarrow} {\lambda}$ are Gell-Mann's
$SU(3)$ matrices with $tr\lambda_a \lambda_b=2\delta_{ab}$)

\be
\Phi_8(x) = {\stackrel{\rightarrow}{\lambda} \over \sqrt 2}.
\stackrel{\rightarrow}{\Phi}_8(x) =
\left( \matrix {{ \pi^0 \over \sqrt 2}+{\eta_8 \over \sqrt 6}&\pi^+&K^+\cr
                         \pi^-&{-\pi^0 \over \sqrt 2}+ {\eta_8
\over \sqrt 6}&K^0\cr
                         K^-&\overline{K}^0&-2{\eta_8
\over \sqrt 6}\cr} \right), \ee

\be
\Phi_1(x)={1\over \sqrt 3} \eta_1 I.
\ee

\vskip 0.4cm

\noindent  The axial-vector field matrix $\xi_{\mu}$ is defined as follows:

\be
\xi_{\mu} = i\{\xi^{\dagger}[\partial_{\mu}-i r_{\mu}]\xi
- \xi[\partial_{\mu}-i l_{\mu}]\xi^{\dagger}\}=i\xi^\dagger D_\mu U
\xi^\dagger, \ee

\vskip 0.4cm

\noindent where $\xi\xi=U$ and $l_\mu$, $r_\mu$ are external $3\times 3$
left and right field matrices. We also define

\be f^{\mu \nu}_{(\pm)} = \xi F_L^{\mu \nu} \xi^{\dagger} \pm
\xi^{\dagger} F_R^{\mu \nu} \xi,   \ee

\vskip 0.4cm

\noindent where $F_{L,R}^{\mu \nu}$ are the external field-strength
tensors

\br\dis
F_L^{\mu \nu}=\partial^\mu l^\nu - \partial^\nu l^\mu - i
[l^\mu,l^\nu],\\\dis
F_R^{\mu \nu}=\partial^\mu r^\nu - \partial^\nu r^\mu - i
[r^\mu,r^\nu].
\er

\vskip 0.4cm

\noindent  Since we are only concerned with $f_{(+)}$ we set in what follows
$f_{(+)}^{\mu\nu}=f^{\mu\nu}$. The specification to the electromagnetic field
reads

\be F_L^{\mu \nu} = F_R^{\mu \nu} = \vert e \vert Q F^{\mu \nu} \ee

\vskip 0.4cm

\noindent where $Q=\hbox{diag}(2/3,-1/3,-1/3)$ is the quark charge matrix.
Finally we shall need

\be
\chi^+=\xi^\dagger\chi\xi^\dagger+\xi\chi^\dagger\xi,
\ee

\vskip 0.4cm

\noindent with

\be
\chi= 2B_0{\cal M},
\ee

\vskip 0.4cm

\noindent where ${\cal M}=\hbox{diag}(m_u,m_d,m_s)$ and $B_0$ is related
to the vacuum
expectation value

\be
\langle \bar q q \rangle = - f_0^2  B_0(1+O({\cal M})).
\ee

\vskip 0.4cm

The structure of the strong chiral Lagrangian up to $O(p^4)$ has been studied
in
\cite{gl}. It is straightforward to write the relevant $O(p^6)$ chiral
Lagrangian
involving two neutral pseudoscalar mesons and two photons.
By restricting ourselves to the operators whose coupling constants are leading
in the large-$N_c$ limit we can write

\be\label{L6}  \dis {\cal L}^{(6)}={2\over f_\pi^2}\left\{
d_1\hbox{tr}\left( \xi^\alpha \xi_\beta f_{\alpha \mu}
f^{\beta \mu}
\right) + d_2\hbox{tr}\left( \xi_\mu \xi^\mu
f_{\alpha \beta} f^{\alpha \beta} \right) + d_3\hbox{tr}\left( \chi^+
f_{\alpha\beta} f^{\alpha \beta}\right) \right\}. \ee

\vskip 0.4cm

\noindent  We have considered here that all quantities were communting since we
are interested only in neutral particles. Our conventions and notation are
chosen in such a way that the expansion of this Lagrangian in terms of
pseudoscalar fields coincides with the analogous expansion of the Lagrangian
(4.28) in
\cite{bgs} which was introduced for the $SU(2)_L\times
SU(2)_R$ case.

In the following section we display how the physical amplitudes are
expressed in terms of these three coupling constants $d_1$, $d_2$ and $d_3$.

\section{\bf Amplitudes in $\chi$PT}

The Lorentz covariant and gauge invariant amplitudes for
$\gamma(q_1)\gamma(q_2)\to \pi^0 (p_1)\pi^0 (p_2)$ and
$\eta(p_1) \to \pi^0(p_2) \gamma(q_1) \gamma(q_2)$ read

\be
A=e^2\epsilon_\mu(q_1)
\epsilon_\nu(q_2)V_{\mu\nu},
\ee

\vskip 0.4cm

\noindent where

\br\dis\label{vmn}
V_{\mu\nu}= A(s,t,u)T_{1\mu\nu}+B(s,t,u)T_{2\mu\nu},\\\dis
T_{1\mu\nu}={s\over 2}g_{\mu\nu}-q_{1\nu}q_{2\mu},\\\dis
T_{2\mu\nu}=2s\Delta_\mu\Delta_\nu-(t-u)^2g_{\mu\nu}-2(t-u)(q_{1\nu}\Delta_\mu-
q_{2\mu}\Delta_\nu),\\\dis
\Delta_\mu=(p_1-p_2)_\mu.
\er

\subsection{\bf $\gamma \gamma \to \pi^0 \pi^0$}


For the $\gamma \gamma \to \pi^0 \pi^0$
amplitude the Lagrangian (\ref{L6}) generates the $O(p^6)$ counterterms

\br\dis
A_6^N={20\over 9 f_\pi^4}[16(d_3^r-d_2^r)m_\pi^2+(d_1^r+8d_2^r)s]+\cdots,
\\\dis
B_6^N=-{10\over 9 f_\pi^4}d_1^r +\cdots.
\er

\vskip 0.4cm

\noindent We refer to Ref. \cite{bgs} for the calculation in $SU(2)_L\times
SU(2)_R$ of the additional contributions coming from chiral loops indicated
here by the ellipses. The contribution of one kaon loop in the $O(p^4)$
amplitude has been computed in \cite{bico}. For $s$ far below the
$K{\bar K}$ threshold it is numerically small, with respect
to the contribution of one pion loop,
owing to an extra factor $\frac{s}{48m_K^2}$
\cite{dhli,bhn}. The calculation of the two-loop amplitude
in $SU(3)_L\times SU(3)_R$ has not been carried out.

As in \cite{bgs} one connects the constants $a_1^r$, $a_2^r$ and
$b^r$ parametrizing the renormalized amplitudes
\br\dis
A_6^N={a_1^r m_\pi^2+ a_2^r s\over (16\pi^2 f_\pi^2)^2}+\cdots,\\\dis
B_6^N={b^r\over (16\pi^2 f_\pi^2)^2}+\cdots, \label{bb1}
\er

\vskip 0.4cm

\noindent with the constants $d_i^r$

\br
\dis {a_1^r\over (16 \pi^2 f_\pi^2)^2}  =
\left(20\over 9\, f_\pi^4\right) 16 (d_3^r - d_2^r),\\\dis
{a_2^r\over (16 \pi^2 f_\pi^2)^2} =
\left(20\over 9\, f_\pi^4\right) (d_1^r + 8\,d_2^r),\\\dis
{b^r\over (16 \pi^2
f_\pi^2)^2} = -\left(10\over 9\, f_\pi^4\right)  d_1^r.
\er

\subsection{\bf $\eta \to \pi^0 \gamma \gamma$}

For the $\eta$ decay we have \cite{abbc}

\br\dis
A_6^\eta={4\sqrt 2\over 3 \sqrt 2 f_\pi^4}[16(d_3^r m_\pi^2-d_2^r m_\eta^2)
+(d_1^r+8d_2^r)s]+\cdots,\\\dis
B_6^\eta=-{2\sqrt 2\over 3 \sqrt 2 f_\pi^4}d_1^r +\cdots. \er

\vskip 0.4cm

\noindent In these expressions we have included the contribution of $d_3$ which
has not been taken into account in \cite{abbc}. However the size of this
contribution is small because $m_\pi^2<<m_\eta^2$. The mixing
$\eta-\eta^\prime$ has been treated within $\chi$PT in \cite{gl,dhl}. As
a result the physical $\eta$ is a superposition

\be
\eta=\cos \theta \eta_8 - \sin \theta \eta_1,
\ee

\vskip 0.4cm

\noindent with $\sin \theta \simeq -{1\over 3}$.

For the expressions of the $O(p^4)$ one-loop contribution to this process we
refer to Ref. \cite{abbc}. As explained there, this contribution is very small
because pion loops violate G-parity and kaon loops are suppressed by a factor
$1/24m_K^2$. The two-loop calculation has never been performed but the same
argument can be advocated \cite{abbc}, to claim that loop contributions to
the $O(p^6)$ amplitude are suppressed. In addition, consistency of the $\chi$PT
expansion implies that they are smaller than the $O(p^4)$ one-loop
contribution.

On the other hand at $O(p^8)$ a new type of one-loop contribution appears  by
taking two anomalous vertices of $O(p^4)$. As shown in \cite{abbc} the latter
has an order of magnitude comparable with the $O(p^4)$ one-loop contribution.
This fact does not break the perturbative $\chi$PT expansion because the higher
order corrections to this $O(p^8)$ one-loop term will be small with respect to
it.

Because all these loop contributions are small with respect to the measured
decay width, one should expect that the $O(p^6)$ counterterms account for a
large part of the full amplitude.

\subsection{\bf Neutral pion polarizabilities}

The polarizabilities caracterize the electric and magnetic properties of a
composite system. They appear as parameters in the low-energy expansion of the
Compton amplitudes at threshold \cite{revpol}

\be
T^{\small\hbox{Compton}} =
2 \left[ \vec{\epsilon}_1 \cdot \vec{\epsilon}_2 \,\! ^\star
 \left(
         {\alpha\over m_\pi} - {\alpha}^N \omega_1 \omega_2
         \right)- {\beta}^N \left(\vec{q}_1 \times \vec{\epsilon}_1
         \right) \cdot \left(
         \vec{q}_2 \times \vec{\epsilon}_2 \,\! ^\star \right)
          + \cdots \right] \co
\ee

\vskip 0.4cm

\noindent where $q_i^\mu=(\omega_i,\vec{q}_i)$. Following the Condon-Shortley
phase convention we define

\br\dis
(\alpha+\beta)^N= 8\alpha m_\pi \lim_{s\to 0} \lim_{t\to m_\pi^2}B,\\
\dis
(\alpha-\beta)^N={\alpha\over  m_\pi} \lim_{s\to 0} \lim_{t\to m_\pi^2}
(A+8m_\pi^2 B),\label{pol}
\er

\vskip 0.4cm

\noindent The one-loop amplitude calculated in \cite{bico,dhli}
has been used in Eqs. (\ref{pol}), in order to find
the neutral pion polarizabilities to $O(p^4)$,
as discussed in Refs. \cite{dhpol,bbgm0}.
Taking into account  the full $O(p^6)$ result one gets
\cite{bgs}

\br\dis
(\alpha-\beta)^N= -{\alpha\over 48 \pi^2 m_\pi f_\pi^2}-
\alpha m_\pi{80 \over 9f_\pi^4}
(d_1^r+ 4 d_2^r - 4 d_3^r) +(\alpha-\beta)^N_{\small 2-\hbox{loop}} ,\\
\dis
(\alpha+\beta)^N= -\alpha m_\pi{80 \over 9f_\pi^4}d_1^r
+(\alpha+\beta)^N_{\small 2-\hbox{loop}},
\er

\vskip 0.4cm

\noindent  where the two-loop contributions can be obtained from Table 3 in
\cite{bgs}\footnote{The values of the
polarizabilities are in units of $10^{-4}\mbox{fm}^3$ in what follows.}

\br\dis
(\alpha-\beta)^N_{\small 2-\hbox{loop}}\simeq -0.31,\\
\dis
(\alpha+\beta)^N_{\small 2-\hbox{loop}}\simeq 0.17,
\er
The latter numerical values have been derived for the $SU(2)_L\times
SU(2)_R$ case; we are neglecting the additional two-loop
contributions for the $SU(3)_L\times
SU(3)_R$ case due to kaon loops. Notice that
the kaon loop contribution to the
one-loop amplitude, calculated in \cite{bico}, vanishes for $s\to 0$. Hence,
at the one-loop level, there is no kaon loop contribution
to the pion polarizabilities, according to (\ref{pol}).

\section{\bf ENJL model prediction for the coupling constants}

The idea of the ENJL model is to approximate large-$N_c$ QCD at the chiral
symmetry breaking scale $\Lambda_\chi$ by an effective four-fermion theory
(generated by integration over quark and gluon fields above
$\Lambda_\chi$). The main assumption of this model is that higher
dimension fermionic operators are irrelevant for long-distances. The three
parameters are the scale $\Lambda_\chi$ and the two coupling
constants
$G_S$ and $G_V$ of the four-fermion operators, respectively the
scalar-pseudoscalar and the vector-axial ones. Alternatively one can trade
these parameters for three other ones, {\it i.e.} the constituent
quark mass,
$M_Q$, the coupling of the constituent quarks to the pseudoscalar current,
$g_A$, and the ratio $M_Q^2/\Lambda_\chi^2$.

Integration over
quarks\footnote{Here we disregard the gluonic fluctuations
below $\Lambda_\chi$ (see \cite{bbr} for a discussion of this
issue).} below $\Lambda_\chi$ (see \cite{bbr} for the
description of the method) yields the low-energy
effective Lagrangians involving pseudoscalar mesons and the lowest spin-0
and spin-1
resonance.
Within this context twenty-two low-energy constants have been derived in
\cite{bbr}
and thirty-six others in \cite{prades}. Whenever there are experimental
data, the ENJL
prediction is in good agreement with them.

A calculation of the coupling constants $d_1$ and $d_2$ in the context of the
chiral quark model which is nothing but the mean field approximation of the
ENJL model, has been carried out in Ref. \cite{bdv} (in this article however
the
operator modulated by $d_3$ in Eq. (\ref{L6}) has not been taken into
account). In this mean field approximation one neglects the fluctuations of the
resonances so that the calculation involves just one loop of a constituent
quark of mass $M_Q$.

In the full ENJL model (see Ref. \cite{bbr}) the quark-loop contribution is
modified with respect to the mean-field approximation by the presence of the
mixing axial-pseudoscalar parametrized by the constant $g_A$. In addition
one gets contribution from integrating out the resonance fields.  For any
coupling our notation is
(in what follows we shall drop the index `r' of the coupling constants)
$d_i ={\tilde d}_i + {\displaystyle \sum^{}_R} d_i^R$ where
${\tilde d}_i$ is the contribution of the quark loop and
${\displaystyle \sum^{}_R} d_i^R$ stands for the sum of the contributions
of the resonances ($R=S,V,A,T$).

Calculations have been performed with the Seeley-DeWitt expansion
\cite{sd1,sd2}. For
details of the procedure the reader can consult Ref. \cite{bbr}.
A brief summary
is given in appendix.

For all the numerical applications we shall use for the input
parameters the values obtained in \cite{bbr}, Fit 1, {\it i.e.} $M_Q = 265$
MeV, $g_A=0.61$ and $x={M_Q^2\over \Lambda_\chi^2}=0.052$.

\subsection{\bf Quark-loop contribution}

The results read

\br\label{denjl}
\dis \vspace*{0.3cm}\tilde d_1=  - {N_c \over 16 \pi^2} {f_\pi^2\over M_Q^2}
{1\over 24}\Gamma_2 \,g_A^2 \co \\\dis
\vspace*{0.3cm}
\tilde d_2={N_c \over 16 \pi^2} {f_\pi^2\over M_Q^2}
{1\over 48}\Gamma_2 \,g_A^2 \co \\\dis
\vspace*{0.3cm}
\tilde d_3={N_c \over 16 \pi^2} {f_\pi^2\over M_Q^2}{1\over 48}\Gamma_1 \,\rho
\,\,\,\,\,\,\,\,\hbox{with} \,\,\,\,\,\,\,\,
\rho = {M_Q f_\pi^2 \over \vert\langle \bar q q \rangle\vert} =
{\Gamma_0\over \Gamma_{-1}}\, g_A \co \er

\vskip 1cm

\noindent where we have used the shortened notation $\Gamma_n=\Gamma(n,M_Q^2/
\Lambda_\chi^2)$ for the incomplete Gamma functions defined as follows:

\be\dis
\Gamma(n-2,x=M_Q^2/\Lambda_\chi^2)=\int_{M_Q^2/\Lambda_\chi^2}^\infty
{d\,z\over z}e^{-z}z^{n-2} \per \label{gamma}
\ee

\vskip 0.4cm

For applications we use $\Gamma_1=\Gamma_2=1$. When one sets
$g_A=1$ our results for $\tilde d_{1,2}$ coincide with those of the quark-loop
calculation of \cite{bdv}.

\subsection{\bf Spin-1 and spin-2 resonances}\label{vect}

Here we shall follow the notations and definitions of
\cite{eglpr,bbr,prades}.
The vector (respectively axial-vector) resonance field will be denoted
by $V_\mu$ (respectively $A_\mu$). We also define

\bd
R_{\mu\nu}=d_\mu R_\nu-d_\nu R_\mu,\,\,\,\,\,\,\,\,\,\,\,\hbox{for}
\,\,\,\,\,\,\,\,\,\,\,R=V,A
\,\,\,\,\,\,\,\,\hbox{with}
\,\,\,\,\,\,\,\,\,\,\,d_\mu=\partial_\mu+\Gamma_\mu \co
\ed

\vskip 0.4cm

\noindent   where $\Gamma_{\mu}$ is given by the expression

\be\dis
\Gamma_\mu={1\over 2}\{\xi^{\dagger}[\partial_{\mu}-i r_{\mu}]\xi
- \xi[\partial_{\mu}-i l_{\mu}]\xi^{\dagger}\} \per
\ee

Let us first consider the vector case. The coupling we are
interested in is the one which modulates the operator
$\epsilon^{\mu\nu\rho\sigma}$tr$\left( V_\mu\{\xi_\nu,f_{\rho\sigma}\}\right)$
called $h_V$. The $O(p^6)$ counterterms receive contributions from
the exchange of $\omega^0$, $\rho^0$ and $\phi$ mesons.
The last one will not be considered here
since its contribution is suppressed by the large $\phi$ mass (the
$\phi$ resonance contribution to the $\gamma\gamma\to\pi^0\pi^0$
amplitude has been included in Refs. \cite{bbho,bgs}).
Guided by the nonet
assumption, we shall not use for the contribution of the $\rho^0$ meson the
experimental data on the decay $\rho^0\to \pi^0 \gamma$ which have a large
error, but instead those on the decay $\rho^+\to \pi^+ \gamma$. Then from
$\Gamma(\rho^+\to \pi^+ \gamma)$ and $\Gamma(\omega^0\to \pi \gamma)$
\cite{pdb} one can extract

\be\label{hv}
{h_{\rho^+}\over M_{\rho^+}}
\simeq 4.7\cdot10^{-5}\,\hbox{MeV}^{-1}\,\,\,\,\,\,\,\,
\hbox{and}\,\,\,\,\,\,\,\,
{h_{\omega^0}\over M_{\omega^0}}\simeq 4.9\cdot10^{-5}\,\hbox{MeV}^{-1}
\ee

\vskip 0.4cm

\noindent and take, according to the nonet assumption,
${h_{\rho^0}\over M_{\rho^0}}\equiv {h_{\rho^+}\over M_{\rho^+}}$.

Within the ENJL model the constant $h_V$ has been computed in \cite{prades}.
Here we just recall the result

\be h_V = {N_c\over 16 \pi^2} {\sqrt{2}\over 8 f_V} (1+g_A), \ee

\vskip 0.4cm

\noindent where $f_V$ modulates the operator
tr$\left(f_{\mu\nu}V^{\mu\nu}\right)$ and has the following expression
\cite{bbr}

\be
f_V^2 = {N_c\over 16 \pi^2} {2\over 3}\Gamma_0\cdot
\ee

\vskip 0.4cm

\noindent Using the relation found in the ENJL model \cite{bbr}

\be
f_V M_V = {f_\pi\over \sqrt{1-g_A}}
\ee

\noindent one obtains

\be {h_V\over M_V} = {N_c\over 16 \pi^2} {\sqrt{2}\over 8 f_\pi}
(1+g_A)\sqrt{1-g_A}\,\cdot \ee

\vskip 0.4cm

\noindent Numerically one gets, for $g_A=0.61$,

\be
{h_V\over M_V}=3.6\cdot 10^{-5}\,\hbox{MeV}^{-1}.
\ee

\vskip 0.4cm

\noindent For this constant the agreement of the ENJL prediction with
experiment (\ref{hv}) is not as impressive as for the other low-energy
constants. Moreover this quantity has to be squared in the contribution
to the amplitude, so that finally the ENJL model prediction of $d_i^V$
is a factor two smaller than the one based on resonance saturation.
However one has to keep in mind that the coupling constants
predicted in the ENJL model are parameters of the Green functions
evaluated {\it at zero momenta}. This fact has to be taken into
account, when comparing with processes, such as the decay of a vector
resonance, whose typical energy scale is large. In \cite{prades} it was argued
that including intermediate resonance exchanges and chiral loops one obtains an
improved prediction in good agreement with the phenomenology of the vector
resonance decay.

Whenever we deal with a processes at small energy, such as Compton scattering
at threshold for the determination of the pion polarizabilities, we favour
the value of $h_V$ predicted in the ENJL model.
On the other hand, for the $\eta$ decay the energy scale is
quite close to the $\rho$ mass and it seems appropriate to use for $h_V$ the
value
extracted from experiments. Finally for
the $\gamma \gamma \to \pi^0 \pi^0$ transition we shall display two different
cross-sections
corresponding to the values of $h_V$ obtained from ENJL and phenomenology,
respectively.
We take the difference
between the cross-sections as a contribution to the uncertainty
on our result.

The $b_1$ axial-vector resonance is not included in the ENJL model. We shall
use
the experimental data in order to incorporate it as it is done in \cite{bgs}.

The tensor resonance is not described either by the ENJL model. Such a
resonance would appear only if higher derivative four-quark operators were
taken into account. In this case again we shall use the experimental data.
Unfortunately they allow only for a determination of the absolute value of the
tensor contribution to the transitions considered here. Hence this contribution
will appear as an uncertainty in our results.

\subsection{\bf Scalar resonance}\label{scal}

The scalar-sector analysis requires the knowledge of the coupling constants
$C_S^d$, $C_S^m$ and $C_S^\gamma$ which modulate respectively the operators
tr$\left( S \xi_\mu \xi^\mu\right)$, tr$\left( S \chi^+\right)$ and
$e^2 F_{\mu\nu} F^{\mu\nu}$tr$\left( S Q^2\right)$.
The coupling constants $C_S^d$, $C_S^m$ as well as the scalar mass $M_S$
have been computed within the ENJL model in \cite{bbr}. Their expressions
reads

\be C_S^d = {N_c\over 16 \pi^2} {M_Q \over \lambda_S}\, 2 g_A^2\,
(\Gamma_0 - \Gamma_1), \ee

\be C_S^m = {N_c\over 16 \pi^2} {M_Q \over \lambda_S}\, \rho\,
(\Gamma_{-1} - 2 \Gamma_0), \ee

\vskip 0.4cm

\noindent with the rescaling factor

\be \lambda_S^2 = {N_c\over 16 \pi^2} {2\over 3} (3\Gamma_0 - 2\Gamma_1)
\ee

\vskip 0.4cm

\noindent  and the mass

\be M_S^2 = {N_c\over 16 \pi^2} {8 M_Q^2 \over \lambda_S^2} \Gamma_0.
\ee

\vskip 0.4cm

\noindent Numerical evaluations for these constants can be found in
\cite{bbr} as well as the comparison with  experiment and with the resonance
saturation approach.

We have computed the coupling constant $C_S^{\gamma}$
which governs the decay $S \to \gamma \gamma$ with the result

\be\label{csg} C_S^{\gamma} = {N_c\over 16 \pi^2} {2\over 3}
{\Gamma_1\over M_Q \lambda_S} \, .  \ee

\vskip 0.4cm

\noindent  Numerically we find $C_S^{\gamma}\simeq 0.19 \,\,\hbox{GeV}^{-1}$.
This value is about a factor 2 larger than the estimate given in \cite{bgs}
(N.B. with an uncertainty of $100\%$). However it corresponds rather
well to the estimate obtained in the analysis of \cite{ms}, where higher
order terms in the chiral expansion are taken into account, in order to match
the QCD high energy behaviour.

The contributions of the scalar particle to the constants $d_i$ (or
$a_{1,2}$ and $b$) are given in \cite{bgs} in terms of the mass $M_S$ and the
absolute values of the constants $C_S^d$, $C_S^m$, $C_S^{\gamma}$
(in contrast with the resonance saturation calculation, here we know the sign
of the different contributions). Within the ENJL model we find

\be
d_1^S = 0,
\ee

\be
d_2^S = {C_S^\gamma C_S^d f_\pi^2\over 8M_S^2}=
\left({N_c \over 16 \pi^2} {f_\pi^2\over M_Q^2}\right)
{1\over 48} {\Gamma_1\over
\Gamma_0}\,g_A^2\,(\Gamma_0-\Gamma_1),
\ee

\be
d_3^S = {C_S^\gamma C_S^m f_\pi^2\over 8M_S^2}=
\left({N_c \over 16 \pi^2} {f_\pi^2\over M_Q^2}\right)
{1\over 96} {\Gamma_1\over
\Gamma_0}\rho(\Gamma_{-1}-2\Gamma_0).
\ee

\vskip 0.4cm

\noindent We find $a_1^S\simeq 0.14$ and
$a_2^S\simeq 4.7$ whereas the resonance saturation approach used
in Ref. \cite{bgs} gives $a_1^S\simeq \pm 0.8$ and
$a_2^S\simeq \pm 1.3$.

A few comments are in order. First we have a definite
prediction for the signs, even though it is very dependent on the input
parameters in the case of $a_1^S$ which is proportional to the difference of
$d_3^S$ and $d_2^S$, two quantities of comparable order of magnitude.
Secondly and as a consequence of the latter remark we confirm the smallness of
$a_1^S$. Finally we get for $a_2^S$ (which is proportional to $d_2^S$) a
sensibly higher value. This is related to the fact that in the ENJL model the
constant $C_S^\gamma$ computed in Eq. (\ref{csg}) appears to be higher than the
estimate of \cite{bgs} (but, as we said, this is in agreement with the recent
analysis of \cite{ms}) and the scalar mass rather low
(see \cite{bbr,brz}).
At the present time it is not completely clear if the scalar
contribution to the coupling constants predicted by the ENJL model is a
better estimate than the analogous contribution calculated in \cite{bgs}.
Hence we keep the latter as a lower bound estimate and obtain in this way
the uncertainty on the scalar contribution to the coupling constants.

In the next section we discuss a possible way of improving the
situation in the future which is related to the coupling constant $d_3$.

\subsection{\bf On the constant $d_3$}

An interesting feature of the scalar sector is that within the
ENJL model many cancellations occur between the quark-loop and the scalar
resonance contributions. This is the case at $O(p^4)$ for the constants
$L_5$, $L_8$ and
$H_2$ (see \cite{bbr}). In our calculation, one finds that the combination
$a_1^S+\tilde a_1$ is actually independent of the parameter $\rho$ and of the
quadratically divergent incomplete gamma function $\Gamma_{-1}$. For the full
$d_3$ one obtains a quite simple expression

\be \label{d3}
d_3 = d_3^S + \tilde d_3 =
\left({N_c \over 16 \pi^2} {f_\pi^2\over M_Q^2}\right)
{1\over 96}g_A\,\Gamma_1.
\ee

\vskip 0.4cm

\noindent Let us also notice that we get the following
relation\footnote{which is valid to all orders in the $\alpha_S N_c$-expansion
(see Ref. \cite{bbr})}:

\be
{C_S^\gamma \over C_S^d} = 8\, {d_3\over f_\pi^2 L_5}.
\ee

\vskip 0.4cm

\noindent With the set of parameters we have already used one finds from Eq.
(\ref{d3}) $d_3\simeq 1.5\times 10^{-5}$ whereas the value predicted in
\cite{bgs} using scalar resonance saturation is
$d_3\simeq 0.38 \times 10^{-5}$.
The difference between these values comes both from $d_3^S$ (which is already
bigger in our case than the value of \cite{bgs}) and from $\tilde
d_3$ wich is not present in the resonance saturation approach. One can compare
both values with the estimate $d_3= (0.94\pm 0.47) \times 10^{-5}$ obtained in
\cite{kms} using a sum rule for the vector-vector two-point function. This
estimate is
affected by a large uncertainty wich makes it compatible with both our value
and the value in \cite{bgs} and so does not provide a very satisfactory test.

An important improvement would be to extend this sum rule to two-loop order, as
it is required by the fact that $d_3$ modulates an $O(p^6)$ operator. Hence it
is a bit
premature to try to improve the agreement of our result with the sum-rule,
which could be achieved by changing slightly the value of the parameters $M_Q$
and $g_A$ and redoing the fits of Ref. \cite{bbr} including $d_3$;
this would of course have an effect on all the $O(p^6)$ constants. We postpone
such considerations until a full two-loop calculation of the sum-rule is
carried
out.

\section{\bf Results}

\subsection{\bf $\gamma \gamma \to \pi^0 \pi^0$}

In the Table we display the full ENJL results for the constants $a_1$, $a_2$
and
$b$ to be compared with Table 2 in \cite{bgs}.

\begin{table}[htb]

\caption{Final results}

\label{table3}
\vspace*{0.4cm}
\begin{center}

\begin{tabular}{|r||r||r|r||r||r|r||r||r|}  \hline
&Q.L.&\multicolumn{2}{|c||}{$(1^{--})$} &$(1^{+-})$
&\multicolumn{2}{|c||}{$(0^{++})$} &$(2^{++})$ &
Total$\;\;\;\;\;\;\;\;\;$ \\ \hline
&&ENJL&Data&&ENJL&Data&& \\ \hline

$a_1$&$-12.3$&$-20.3$&$-36.6$&$0$ &$0.14$&$0.8$&$\mp
4.1 $& $-28 \ge a_1 \ge -53$ \\
\hline
$a_2$&$6.1$ & $7.6$&$13.7$&$-1.3$ &$4.7$&$1.3$ &$\pm 1.0 $&
$13 \le a_2 \le 24\;\;\;$ \\
\hline
$b$&$1.0$ & $1.3$&$2.3$& $0.7$ &$0$&$0$ &$\pm 0.5$ &
$2 \le b \le 4\;\;\;\;\;\;$ \\
\hline
\end{tabular}
\end{center}

\end{table}

\vspace*{0.4cm}

\noindent Inspection of the Table shows that we have three sources of
uncertainties.
The first two are related to the different estimates of the vector and scalar
coupling constants obtained in section 4 by using ENJL and experimental data,
respectively\footnote{We assume the central value of the experimental data,
ignoring the errors on the measured parameters (which typically have a $10\%$
size). In addition, we disregard the changes in
the ENJL prediction due to variations of the three parameters of the model.
In our opinion, assuming the quite large band of error on the cross-section
associated to the ENJL prediction and the experimental estimate
for the resonance contributions, is pessimistic enough.}.
In the last column we have added these
uncertainties in such a way as to have an upper bound and a lower bound for
the cross-section. This is because the uncertainties
on the vector (see section 4.2) and the scalar resonance (see section 4.3)
contribution have different origins.
The last uncertainty comes from the sign of the tensor contribution which
is undetermined. We included also this contribution in the total
(last column of the Table).

For the low-energy constants entering the helicity amplitudes we get

\br\dis
-8 \ge h_+=a_1+8b \ge -17 \co \\\dis
\;\;\;\; 8 \le h_s=a_2-2b \le 15 \co \\\dis
\;\;\;\; 2 \le h_-=b \le 4 \per \label{heli}
\er

\vskip 0.4cm

\noindent Our results are compatible with
those quoted in \cite{bgs}:

\br\dis
h_+=-14\pm 5\co\\\dis
h_s=7\pm 3\co\\\dis
h_-=3\pm 1\per\label{heli2}
\er

\noindent In the Figure we display the data from Ref. \cite{crystal}
for the cross-section $\sigma (s$; $\vert$cos$\theta\vert\le Z=0.8)$
as a function
of the center-of-mass (c.m.) energy $E=\sqrt{s}$. The full
(respectively, dashed) curve in the Figure corresponds to the values in
the l.h.s. (respectively, the r.h.s.) of the inequalities
(\ref{heli}). For comparison we show in the Figure, as a dotted
curve, also
the cross-section obtained in \cite{bgs} for the central values of
(\ref{heli2}) (see the full lines in Figs. 5,11 of Ref. \cite{bgs}).
We can also compare our results with Fig. 9 of Ref. \cite{bgs}, where the
uncertainties in the values of (\ref{heli2}) are taken into account
(in Fig. 9 of Ref. \cite{bgs}, however, the contribution of the
integrals $\Delta_{A,B}$
coming from the two-loop box and the acnode diagrams in Fig. 4 of Ref.
\cite{bgs} is neglected, whereas in our Figure the full contribution
is retained). The upper curve in Fig. 9 of Ref. \cite{bgs} disagrees
with our result by more than 20$\%$ for c.m. energies $E \ge$540 MeV
(at $E=540$ MeV the $\Delta_{A,B}$ contribution to the
cross-section is very small, i.e. 1.5$\%$, see Figs. 5,9 of \cite{bgs}).

\subsection{\bf $\eta \to \pi^0 \gamma \gamma$}

As we said in section 3.2 the $O(p^6)$ counterterms
contribution is expected to
account for a large part of the amplitude because the loop contributions are
suppressed. In Ref. \cite{abbc} it is shown that the
resonance saturation approximation is not sufficient
to explain the experimental
decay rate. Indeed if one takes into account only  vector resonance saturation
one gets

\be\label{vmd}
\Gamma(\eta\to\pi^0 \gamma\gamma)\simeq 0.18 \,\hbox{eV}, \label{vr6}
\ee

\vskip 0.4cm

\noindent to be compared with the experimental value \cite{pdb}
$\Gamma(\eta\to\pi^0 \gamma\gamma)\simeq 0.85\pm0.18 \,\hbox{eV}$.
Let us review the calculation of \cite{abbc}.
By keeping the momenta dependence in the vector propagator the authors
of \cite{abbc} made an `all-order' estimate, {\it i.e.} at $O(p^6)$ and
higher, of the counterterms contribution with the result

\be
\Gamma(\eta\to\pi^0 \gamma\gamma)\simeq 0.31 \,\hbox{eV}. \label{vr}
\ee

\vskip 0.4cm

\noindent Then if in addition one takes into account chiral loops (the
$O(p^4)$ ones as well as the $O(p^8)$ `doubly-anomalous' one-loop diagrams)
one reaches the value

\be\label{0.42}
\Gamma(\eta\to\pi^0 \gamma\gamma)=0.42\pm 0.20 \,\hbox{eV},
\ee

\vskip 0.4cm

\noindent where  the error includes scalar and tensor contributions (whose sign
was not known) as well as a $30\%$ error coming from other contributions such
as
one-loop diagrams involving the $O(p^6)$ Lagrangian (\ref{L6}).

Let us now show step by step the ENJL results. To the estimate
(\ref{vmd}) which incoporates only vector resonance exchange, we first add
quark loop contribution and find

\be
\Gamma(\eta\to\pi^0 \gamma\gamma)\simeq 0.36 \,\hbox{eV}.
\ee

\vskip 0.4cm

\noindent Already we are not far from the central value of Eq. (\ref{0.42}).
In addition if we include the scalar contribution as predicted by ENJL we have

\be
\Gamma(\eta\to\pi^0 \gamma\gamma)\simeq 0.5 \,\hbox{eV}.
\ee

\vskip 0.4cm

\noindent We shall use as lower bound the prediction for the scalar exchange
given in \cite{bgs}. Then we have

\be
\Gamma(\eta\to\pi^0 \gamma\gamma)=0.45\pm 0.05 \,\hbox{eV}.
\ee

\vskip 0.4cm

\noindent Finally taking into account the axial resonance and, as in
\cite{abbc}, the contributions from chiral loops and, in addition, an
uncertainty from the contribution of the tensor resonance, we get

\be
\Gamma(\eta\to\pi^0 \gamma\gamma)=0.58\pm 0.12 \,\hbox{eV}.
\ee

\vskip 0.4cm

\noindent Of course we can play the game of making  an `all-order' estimate of
the counterterm given by vector resonance exchange
as it is done in \cite{abbc}.
Starting from the upper bound in the last equation, we obtain a decay rate of
$0.86$ eV. However this represents only a partial resummation of the chiral
corrections. First there are $O(p^8)$
one-loop corrections involving the $O(p^6)$
Lagrangian (\ref{L6}) which could be as important as the
countertems contributions (even though they are next-to-leading in $1/N_c$).
Secondly, in the large $N_c$-limit, other
chiral corrections ({\it i.e.} of $O(p^8)$ and higher) coming from
the counterterms occur within the ENJL model.
Indeed the quark loop contribution considered here corresponds to the
computation of a four-point function at zero momenta.
The full ENJL calculation at non-zero momenta
(as it has been done in \cite{bipr}
for the two-point and some of the three-point
functions) would bring additional chiral
corrections. We shall include all chiral corrections of this type in an
additional $30\%$ uncertainty. Our prediction is then

\be
\Gamma(\eta\to\pi^0 \gamma\gamma)= 0.58\pm 0.3 \,\hbox{eV}.
\ee

\vskip 0.4cm

As we said above, the higher order corrections included in
the `all-order' estimate of the vector resonance contribution
enhance the decay width and therefore go in the right
direction, in order to match the experimental result
$\Gamma(\eta\to\pi^0\gamma\gamma)=0.85\pm 0.18$ eV.
It would be interesting
to carry out the $O(p^8)$ loop analysis
(this can be done, given that
most of the loop diagrams are suppressed (see section 3.2)),
as well as to calculate
in the ENJL model the $O(p^8)$ counterterms. In this way one could
obtain an accurate description of the $O(p^8)$ chiral corrections.

\subsection{\bf Polarizabilities}

The pion polarizabilities have been estimated through
dispersion sum rules \cite{revpol}
\bea
(\alpha - \beta)^N &=& -10 \pm 4 \; \;  \co \nn
\apbn &=& 1.04 \pm 0.07 \; \; \per \label{csp4}
\eea

At order $O(p^6)$ resonance saturation gives \cite{bgs}
\bea
(\alpha -\beta )^N &=&-1.01 - 0.31 -0.58\pm 0.20=-1.90\pm 0.20 \co \nn
(\alpha +\beta )^N &=&0.00 + 0.17 +1.00\pm 0.30=1.17\pm 0.30 \co
\label{polbgs}
\eea
where the three contributions that add up to the final results
on the r.h.s. are the one-loop result, the two-loop contribution, and the
one from the $O(p^6)$ counterterms assuming resonance saturation,
respectively.
There is a large contribution
from the $O(p^6)$ low-energy constants due mainly to
the $\omega$-exchange in the resonance saturation approach
\cite{ba2loop}. For the estimate of the uncertainties
in (\ref{polbgs}) the reader is invited to consult Ref. \cite{bgs}.

The contribution due to the $O(p^6)$ counterterms to the
pion polarizabilities
\bea
(\alpha -\beta )^N_{c.t.} &=& \frac{\alpha m_{\pi}}{(4\pi f_{\pi})^4}
h_+ \co \nn
(\alpha +\beta )^N_{c.t.} &=& \frac{\alpha m_{\pi}}{(4\pi f_{\pi})^4}
8h_- \co \nn
\label{pole}
\eea
is obtained from the ENJL model prediction for the low-energy
constants displayed in Eqs. (\ref{heli})

\br\dis
-0.33 \ge (\alpha -\beta )^N_{c.t.} \ge -0.69 \co \\\dis
\;\;\;\; 0.78 \le (\alpha +\beta )^N_{c.t.} \le 1.44 \per \label{polct}
\er

\vskip 0.4cm

\noindent Adding the one-loop and two-loop contributions - identical
to those in (\ref{polbgs}) - we get our results with the ENJL model

\br\dis
-1.65 \ge (\alpha -\beta )^N \ge -2.01 \co \\\dis
\;\;\;\; 0.95 \le (\alpha +\beta )^N \le 1.61 \co \label{polenjl}
\er

\vskip 0.4cm

\noindent which are compatible with Eqs. (\ref{polbgs}).
Both the prediction in (\ref{polenjl}) and the result in \cite{bgs}
are compatible with the forward
sum rule $\apbn = 1.04 \pm 0.07$ in (\ref{csp4}).

As we mentioned in section 4.2, we consider the ENJL value of
the vector resonance coupling as favoured. This choice
corresponds to the range of values
$-1.65 \ge (\alpha -\beta )^N \ge -1.68$
and $0.95 \le (\alpha +\beta )^N \le 1.28$. Here the residual
uncertainty for $(\alpha -\beta )^N$ reflects an incomplete
knowledge (see section 4.3) in estimating the coupling constants
for the scalar resonance, whereas the undetermined sign of the tensor
contribution to the low-energy constants in the Table
yields the remaining uncertainty on $(\alpha +\beta )^N$.
The sizeable
difference between the value of $(\alpha -\beta )^N$ obtained
using the ENJL prediction for the vector resonance coupling
and the result (\ref{polbgs}) is mainly due
to the fact that the ENJL prediction of the vector contribution
to $h_+$ is a
factor two smaller than the one based on resonance saturation
and the experimental data (see section 4.2).
The size of the quark-loop contribution, which is included in the ENJL
prediction, is not large enough, to bring the ENJL value for
$h_+$ to the one obtained from resonance saturation.

The construction of unitarized $S$-wave amplitudes for
$\gamma \gamma \rightarrow \pi \pi$
 which contain   $\ambcn$ as adjustable parameters
has been carried out in Ref. \cite{kaloshin}. In this case,
only $\ambcn$ can be determined from the data \cite{crystal,dcharged},
with the result \cite{kaloshin}
\bea
  \ambc &=& \; \;\; 4.8 \pm 1.0\; \; \co \nn
 (\alpha - \beta)^N &=& -1.1 \pm 1.7 \; \;  \per
\label{cs5}
\eea
The value (\ref{cs5}) for $(\alpha - \beta)^N$ is consistent with the
two-loop result for the neutral pion, both for the calculation
based on resonance saturation \cite{bgs} and for the ENJL prediction,
whereas the corresponding two-loop
calculation for charged pions is not available and so it cannot
be compared with the value (\ref{cs5}) for $\ambc$.

In \cite{pp} the validity of the errors
quoted in a recent estimate of
$\apbcn$ by Kaloshin and collaborators \cite{kalo} is questioned.
Here the polarizabilities appear as adjustable parameters
in the unitarized D-wave amplitudes, hence the values
of $\apbcn$ can be determined from the data
with the result \cite{kalo}
\bea
  \apbc &=& 0.22 \pm 0.06\; \; \cite{dcharged} \co \nn
 \apbn &=& 1.00 \pm 0.05 \; \;  \cite{crystal} \per
\label{cs55}
\eea
The authors of \cite{pp}, arguing on the partial wave analysis
of the data that shows large uncertainties even
at the $f_2$(1270) mass, conclude that the errors quoted in
(\ref{cs55}) for $\apbn$ are unbelievably small.
This result is compatible with the ENJL prediction (\ref{polenjl}),
as well as with the value in (\ref{polbgs}).

\section{\bf Conclusion}

Within the ENJL model we have computed the $O(p^6)$ coupling constants entering
the $\chi$PT expansion for the amplitudes of $\gamma \gamma \to \pi^0\pi^0$
and $\eta \to \pi^0 \gamma \gamma$. In addition to the contribution
of the resonance exchange, one has to add terms coming from the constituent
quark loop. Both contributions have comparable orders of magnitude.
This situation at $O(p^6)$ differs from the one at $O(p^4)$,
which is described in \cite{bbr}.

Our major result concerns the decay
$\eta \to \pi^0 \gamma \gamma$, where the contribution of the
counterterms dominates the amplitude. The values of the couplings
calculated in the ENJL model yield a prediction for the decay width
which is compatible with the data. Our central value compares
much more favourably with the experimental result \cite{pdb} than the value
calculated in \cite{abbc} within resonance saturation.

Concerning the transition $\gamma \gamma \to \pi^0\pi^0$,
our results are compatible with those of the resonance saturation
approach worked out by \cite{bgs}, as well as with the data \cite{crystal}.
For the neutral pion polarizabilities, we find results consistent
with the estimates given in earlier references \cite{bgs,kaloshin,kalo}.

To improve our knowledge of the coupling constants $d_i$'s and be able
to test in a more accurate way the ENJL prediction,
we need both experimental and theoretical progress.

High-precision data from DA$\Phi$NE
on $\gamma\gamma\rightarrow\pi^0\pi^0$ may allow
to extract the value of $h_s$, since the cross-section
in Fig. 9 of \cite{bgs} shows a sizeable dependence on this low-energy
constant for energies near 600 MeV. It will be necessary to carry out
a unitarization of the two-loop result, using a procedure analogous
to \cite{dohod} and matching the dispersive calculation for the
all-order amplitude with the two-loop amplitudes for
$\gamma\gamma \rightarrow \pi^0 \pi^0$ and
$\gamma\gamma\rightarrow\pi^+\pi^-$, once the latter will have been
calculated. The consideration of this improved, unitarized amplitude
will justify the inclusion in the analysis of experimental
data up to 600 MeV.

For the other source of information, i.e. the
decay $\eta\to\pi^0\gamma\gamma$, it will be
very interesting to investigate the $O(p^8)$ analysis,
as we said in section 5.2.
The crucial test may be the future SATURNE experiments which may allow
to determine $d_1$, $d_2$ as
well as the size of the $O(p^8)$ chiral corrections.


Finally the determination of the constant $d_3$ from the sum-rule has to be
improved by carrying out the computation to the two-loop order.

\vspace{2cm}

\noindent
{\bf Acknowledgement}\\
We are pleased to thank Ll. Ametller, J. Bijnens, A. Bramon,
J. Gasser, J. Prades, E. de Rafael,
M.E. Sainio and J. Stern for useful discussions.
\newpage

\appendix

{\bf \huge Appendix}

\vskip 1cm

Let us define

\bo\dis
\nabla_\mu=\partial_\mu+{\cal A}_\mu \co
\eo

\bo\dis
{\cal A}_\mu=\Gamma_\mu-{i\over 2}\gamma_5 (g_A \xi_\mu + W_\mu^{(-)})
-{i\over 2}W_\mu^{(+)} \co
\eo

Let us also introduce the following expression:

\bo\dis
M=-M_Q-S-{1\over 2}(\Sigma-\gamma_5\Delta) \co
\eo
where
\bo\dis
\Sigma=\xi^\dagger{\cal M}\xi^\dagger+\xi{\cal M}^\dagger\xi \co \\\dis
\Delta=\xi^\dagger{\cal M}\xi^\dagger-\xi{\cal M}^\dagger\xi \per
\eo

\vskip 0.4cm

\noindent Here $S$, $W_\mu^{(+)}$ and $W_\mu^{(-)}$
denote the fields of the
resonances, respectively scalar, vector and axial-vector.

We recall briefly the Seeley-DeWitt expansion of the operator

\bd
{\cal D}_E^{\dagger}{\cal D}_E-M_Q^2=-\nabla_\mu\nabla_\mu+E \co
\ed

\vskip 0.4cm

\noindent where ${\cal D}_E$ is the Euclidean Dirac operator

\bd
{\cal D}_E = \gamma_\mu \nabla_\mu + M
\ed

\vskip 0.4cm

\noindent and

\bo\dis
E= 2M_Q S +iM_Q\gamma_\mu\gamma_5 g_A
\xi_\mu -{1\over 4}\sigma_{\mu\nu}f_{\mu\nu} +\cdots
\eo

\vskip 0.4cm

\noindent (here we do not need the complete expression which can be found in
\cite{bbr}).

The effective action reads then

\bo\dis
\Gamma_E=-{1\over 2} \int_{1/\Lambda_\chi^2}^\infty {d\tau \over \tau}
\hbox{Tr exp}(-\tau {\cal D}_E^{\dagger}{\cal D}_E )=
-{1\over 2} {N_c \over 16\pi^2}\sum_{n=0}^\infty
{\Gamma(n-2,x)\over (M_Q^2)^{n-2}}\hbox{tr}H_n \co
\eo

\vskip 0.4cm

\noindent in terms of the incomplete Gamma functions defined in (\ref{gamma}).
We give the Seeley-Dewitt coefficients up to a total derivative and a circular
permutation. Only the terms we are interested in are displayed here.
These terms read

\vskip 0.4cm

\bo\dis H_3=-{1\over 6}\left\{E^3 +{1\over 2}E R_{\mu\nu}R_{\mu\nu}
-{1\over
2}E\nabla_{\mu}\nabla_{\mu}E +{1\over
10}\nabla_{\mu}R_{\mu\rho}\nabla_{\nu}R_{\nu\rho}\right\} \co \\\\\dis
H_4={1\over 24}\left\{ E^4+{1\over 5}[E R_{\mu\nu}E R_{\mu\nu}+ 4E^2
R_{\mu\nu}R_{\mu\nu}] \right\} \per \eo

\vskip 0.4cm

We have also  made use of the following identity:

\bo\dis
\hbox{tr}\{(d_\mu f_{\nu\rho})^2\}=\hbox{tr}\{2[(d_\mu f_{\mu\nu})^2
+i f_{\mu\nu}f_{\nu\rho}f_{\rho\mu}
+\xi_\mu f_{\nu\rho} f_{\rho\mu}\xi_\nu]\\\dis
+{1\over 2}[\xi_\mu f_{\nu\rho}\xi_\mu f_{\nu\rho}
+\xi_\mu \xi_\mu f_{\nu\rho} f_{\nu\rho}
+\xi_\mu f_{\nu\rho} \xi_\nu f_{\rho\mu}
+\xi_\mu f_{\nu\mu} \xi_\rho f_{\rho\nu}]\} \per
\eo

\end{document}